\documentclass[12pt]{article}
\usepackage{amsmath}
\usepackage{latexsym}
\usepackage{amsfonts}
\usepackage[normalem]{ulem}
\usepackage{array}
\usepackage{amssymb}
\usepackage{graphicx}
\usepackage[backend=biber,
style=numeric,
sorting=none,
isbn=false,
doi=false,
url=false,
]{biblatex}

\usepackage{subfig}
\usepackage{wrapfig}
\usepackage{wasysym}
\usepackage{enumitem}
\usepackage{adjustbox}
\usepackage{ragged2e}
\usepackage[svgnames,table]{xcolor}
\usepackage{tikz}
\usepackage{longtable}
\usepackage{changepage}
\usepackage{setspace}
\usepackage{hhline}
\usepackage{multicol}
\usepackage{tabto}
\usepackage{float}
\usepackage{multirow}
\usepackage{makecell}
\usepackage{fancyhdr}
\usepackage[toc,page]{appendix}
\usepackage[hidelinks]{hyperref}
\usetikzlibrary{shapes.symbols,shapes.geometric,shadows,arrows.meta}
\tikzset{>={Latex[width=1.5mm,length=2mm]}}
\usepackage{flowchart}\usepackage[paperheight=11.0in,paperwidth=8.5in,left=1.0in,right=1.0in,top=1.0in,bottom=1.0in,headheight=1in]{geometry}
\usepackage[utf8]{inputenc}
\usepackage[T1]{fontenc}
\TabPositions{0.5in,1.0in,1.5in,2.0in,2.5in,3.0in,3.5in,4.0in,4.5in,5.0in,5.5in,6.0in,}

\setlistdepth{9}
\renewlist{enumerate}{enumerate}{9}
		\setlist[enumerate,1]{label=\arabic*)}
		\setlist[enumerate,2]{label=\alph*)}
		\setlist[enumerate,3]{label=(\roman*)}
		\setlist[enumerate,4]{label=(\arabic*)}
		\setlist[enumerate,5]{label=(\Alph*)}
		\setlist[enumerate,6]{label=(\Roman*)}
		\setlist[enumerate,7]{label=\arabic*}
		\setlist[enumerate,8]{label=\alph*}
		\setlist[enumerate,9]{label=\roman*}

\renewlist{itemize}{itemize}{9}
		\setlist[itemize]{label=$\cdot$}
		\setlist[itemize,1]{label=\textbullet}
		\setlist[itemize,2]{label=$\circ$}
		\setlist[itemize,3]{label=$\ast$}
		\setlist[itemize,4]{label=$\dagger$}
		\setlist[itemize,5]{label=$\triangleright$}
		\setlist[itemize,6]{label=$\bigstar$}
		\setlist[itemize,7]{label=$\blacklozenge$}
		\setlist[itemize,8]{label=$\prime$}

\begin{document}
\section*{A Description of a Subtask Dataset with Glances}
\addcontentsline{toc}{section}{A Description of a Subtask Dataset with Glances}
B. D. Sawyer\textsuperscript{1}, Sean Seaman\textsuperscript{2}, Linda Angell\textsuperscript{2}, Jon Dobres\textsuperscript{1}, Bruce Mehler\textsuperscript{1}, $\&$  Bryan Reimer\textsuperscript{1}\par

\textsuperscript{1} AgeLab, Massachusetts Institute of Technology, \textsuperscript{2 }Touchstone Evaluations, Inc.\par

\subsection*{Author Note}
\addcontentsline{toc}{subsection}{Author Note}
Address correspondence to Dr. Ben D. Sawyer; bsawyer@mit.edu or sawyer@inhumanfactors.com\par

\subsection*{Abstract}
\addcontentsline{toc}{subsection}{Abstract}
This paper describes a set of data made available that contains detailed subtask coding of interactions with several production vehicle HMIs on open roadways, along with accompanying eyeglance data. \par

\subsection*{Introduction}
\addcontentsline{toc}{subsection}{Introduction}
The purpose of this paper is to describe a dataset prepared specifically for understanding the relationship between the modality demands of subtasks and subtask- and task-level glance behavior. This data is drawn from several studies conducted by MIT’s AgeLab (see Method, below) on open roads in production vehicles. The data were originally manually dual-coded and mediated by video analysts for glance behavior; the data were then coded a second time by video analysts for subtask behavior (and additional glance behavior, when needed). \par

\subsection*{Method}
\addcontentsline{toc}{subsection}{Method}
Each of these studies included in this dataset was conducted on Boston-area highways in six different production vehicles that represented modern HMIs with combinations of visual-manual and voice-based infotainment tasks. These studies are described in depth in Mehler et al. (2016), Reimer et al. (2015), Mehler et al. (2014a), Mehler et al. (2014b), Mehler et al. (2015a), and Mehler et al. (2015b), although in the shared dataset the correspondence between specific study publication and study code have been made arbitrary to protect the privacy of participants. Across all studies, participants ranged in age from 20 to 69 years and were evenly split between male and female (49.4$\%$  of participants were female). Also, across all studies, participants possessed valid driver’s licenses, and were naïve as to the purpose of the studies for which they were recruited. Each study’s primary purpose was to evaluate the visual demand associated with HMI interaction. To assess visual demand, videos were taken of participants’ faces as they performed tasks, as well as each vehicle’s center stack. Audio recordings of HMI interactions were also made. \par

Glance analysts utilized the recordings of participants’ faces to manually code the location to which each participant was looking at any given time. Glances and transitions were coded following ISO 15007 IS (2014), with each glance subtending the duration from the first frame identified of the leading transition to the new glance location, through the fixation at the new location, and ending immediately before the first frame indicating the transition to the next location. The locations included in this dataset are described below in Table 2. Tasks with video that could not be classified were excluded from this dataset.\par

Subtask coding began with the review of 10 participants tasks, using a combination of face video, HMI video, and audio recording, in order to create descriptive $``$typical$"$  interaction between system and operator over the course of the task. This script was then subdivided into subtasks, according to the antiphony framework (Sawyer, Mehler $\&$  Reimer, 2018). For each participant each subtask onset and offset were coded, or the first frame indicating movement toward an interaction through the completion of the interaction. Notably, this period often extended beyond the commonly used time window (e.g., NHTSA, 2013) between a clear cue to $``$begin$"$  and a clear cue the participant is $``$done.$"$  Because of this dataset’s focus on capturing all glance behavior associated with the subtasks used to complete a task, glance coding was done on the entire period subtended between the first subtask and the last subtask, rather than the more common $``$recording period$"$  between a clear $``$begin$"$  cue and a $``$done$"$  cue. Thus, subtask and glance coding were entirely overlapping periods, and after aligning the two datasets, each subtask had associated glance behavior. Tasks where entirely overlapping data were not available were excluded (< 1$\%$  of available data). \par

Each participant had their own unique path through the interface, in terms of subtasks. This was in part due to errors and subsequent necessary corrections, but non-erroneous unique paths were found especially within voice interfaces, where multiple paths often existed to achieve a given task. We also encountered challenges with manual tasks. An individual button press, especially one performed with the sequence of other button presses, may be so brief as to subtended glances, rather than being subtended by glances. The method we undertook to simplify this issue was the grouping of multiple sequential button presses into ‘operation groups’, signified by the same metadata as any other touch operation. This was initially developed to handle phone numbers of varying lengths, but soon proved a useful tool to handle a variety of similar situations. \par

Each\ subtask\ (or step) of a task was characterized by the modalities of attentional resources that it required of the driver. These included touch (for manual interactions), hear (for auditory cues and messages provided by the HMI), speak (for voice-based interactions), and vision.  However, the interface demand placed on the human’s visual resources was handled differently in order to prevent circularity and $``$overlap$"$  with the measures that the model was seeking to predict.  The interface demand on the human’s visual resources was conceptualized as $``$display monitoring$"$  resources -- and coded in terms of what was made available by the vehicle system for display (for visually-presented task-relevant information). Thus, touch, hear, speak, and display monitoring (T,H,S,D) resources were coded in a binary fashion; a $``$0$"$  was used to indicate the absence of this demand for a particular subtask, while a $``$1$"$  was used to indicate the presence. The display demand modality was different; this score represented the number of discrete displays containing task-relevant visual information, and this score ranged from 0 (no relevant displays) to 2 (2 relevant displays). As such, vector representation was available for subtasks subtending classical auditory-vocal interactions (0010), visual-manual interactions (1001), and mixed mode interactions (1010), as well as other combinations of interface requirement. Any time before the infotainment system or the driver engaged in a subtask was coded as $``$latency,$"$  per the antiphony cycle, and was represented with both hear and display monitoring requirements (0101). This represented an assumption that many interfaces are interruptible, or may enter confirmation modes, or even self-terminate, and the driver must monitor for cues to such changes in the structure of the task. These canonical demands for each subtask of a task were only included if latency was observed in the video. \par

\subsection*{Data}
\addcontentsline{toc}{subsection}{Data}
The dataset is made available in JSON format (see $``$Introducing JSON,$"$  (n.d.) at \href{https://www.json.org/}{https://www.json.org/}\ for a description).  Each row contains data about one subtask for one trial of a task, organized by participant, study, and, when more than one vehicle is present in a study, by vehicle. Fields associated with each row are described in Table 1. The first glance indicated for each subtask is the location to which the participant was looking when the subtask was coded as beginning. \par

\vspace{\baselineskip}

\vspace{\baselineskip}

\vspace{\baselineskip}
\par


{
\setlength\extrarowheight{3pt}
\begin{longtable}{p{1.19in}p{3.98in}}
\caption{. Glance locations and descriptions.}\label{tab:. Glance locations and descriptions.}

\endfirsthead
\multicolumn{2}{c}{\textit{continued from previous page}}\\
\hline
\endhead\hline
\multicolumn{2}{r}{\textit{continued on next page}} \\
\endfoot
\hline 
\endlastfoot\hline
\multicolumn{1}{|p{1.19in}}{Code} & 
\multicolumn{1}{|p{3.98in}|}{Definition} \\
\hhline{--}
\multicolumn{1}{|p{1.19in}}{Study} & 
\multicolumn{1}{|p{3.98in}|}{String: uniquely (but arbitrarily) identifies each study} \\
\hhline{--}
\multicolumn{1}{|p{1.19in}}{Participant} & 
\multicolumn{1}{|p{3.98in}|}{Integer: uniquely (but arbitrarily) identifies each participant within the dataset} \\
\hhline{--}
\multicolumn{1}{|p{1.19in}}{TaskCode} & 
\multicolumn{1}{|p{3.98in}|}{String: uniquely identifies each task within a study} \\
\hhline{~~}
\multicolumn{1}{|p{1.19in}}{Trial} & 
\multicolumn{1}{|p{3.98in}|}{Integer: ordinal identifier of each repetition of each task (this value is typically 1, but occasionally more than 1 trial is present)} \\
\hhline{~~}
\multicolumn{1}{|p{1.19in}}{StartTime} & 
\multicolumn{1}{|p{3.98in}|}{Decimal: start time of subtask, in seconds; useful for verifying temporal order of subtasks within a task.} \\
\hhline{--}
\multicolumn{1}{|p{1.19in}}{Vehicle} & 
\multicolumn{1}{|p{3.98in}|}{String: uniquely (but arbitrarily) identifies each vehicle} \\
\hhline{--}
\multicolumn{1}{|p{1.19in}}{Gender} & 
\multicolumn{1}{|p{3.98in}|}{String: gender of participant (M or F)} \\
\hhline{--}
\multicolumn{1}{|p{1.19in}}{Age} & 
\multicolumn{1}{|p{3.98in}|}{Integer: age of participant, in years} \\
\hhline{--}
\multicolumn{1}{|p{1.19in}}{Display} & 
\multicolumn{1}{|p{3.98in}|}{Integer: display metadata code} \\
\hhline{--}
\multicolumn{1}{|p{1.19in}}{Touch} & 
\multicolumn{1}{|p{3.98in}|}{Integer: touch metadata code} \\
\hhline{--}
\multicolumn{1}{|p{1.19in}}{Hear} & 
\multicolumn{1}{|p{3.98in}|}{Integer: hear metadata code} \\
\hhline{--}
\multicolumn{1}{|p{1.19in}}{Speak} & 
\multicolumn{1}{|p{3.98in}|}{Integer: speak metadata code} \\
\hhline{--}
\multicolumn{1}{|p{1.19in}}{Latency} & 
\multicolumn{1}{|p{3.98in}|}{Integer: latency metadata code} \\
\hhline{--}
\multicolumn{1}{|p{1.19in}}{SubtaskDuration} & 
\multicolumn{1}{|p{3.98in}|}{Decimal: length of trial of subtask, in seconds} \\
\hhline{--}
\multicolumn{1}{|p{1.19in}}{GlanceStartSeconds} & 
\multicolumn{1}{|p{3.98in}|}{Array of decimals: start time of each glance away from the road} \\
\hhline{--}
\multicolumn{1}{|p{1.19in}}{GlanceEndSeconds} & 
\multicolumn{1}{|p{3.98in}|}{Array of decimals: end time of each glance away from the road} \\
\hhline{--}
\multicolumn{1}{|p{1.19in}}{GlanceLocation} & 
\multicolumn{1}{|p{3.98in}|}{Array of strings: locations of each glance (see Table 2)} \\
\hhline{--}

\end{longtable}}


\vspace{\baselineskip}
Glance locations for glances included in this dataset are described in Table 2.\par

\par


\begin{table}[H]
 			\centering
\begin{tabular}{p{1.18in}p{3.36in}}
\hline
\multicolumn{1}{|p{1.18in}}{Glance location} & 
\multicolumn{1}{|p{3.36in}|}{Description} \\
\hhline{--}
\multicolumn{1}{|p{1.18in}}{road} & 
\multicolumn{1}{|p{3.36in}|}{The forward windshield} \\
\hhline{~~}
\multicolumn{1}{|p{1.18in}}{center stack} & 
\multicolumn{1}{|p{3.36in}|}{The center stack of each vehicle} \\
\hhline{~~}
\multicolumn{1}{|p{1.18in}}{left} & 
\multicolumn{1}{|p{3.36in}|}{Left window and/or left side mirror} \\
\hhline{~~}
\multicolumn{1}{|p{1.18in}}{right} & 
\multicolumn{1}{|p{3.36in}|}{Right window and/or right side mirror} \\
\hhline{~~}
\multicolumn{1}{|p{1.18in}}{rearview mirror} & 
\multicolumn{1}{|p{3.36in}|}{Rearview mirror mounted on or near windshield} \\
\hhline{~~}
\multicolumn{1}{|p{1.18in}}{instrument cluster} & 
\multicolumn{1}{|p{3.36in}|}{Instrument cluster (with speedometer, etc.) } \\
\hhline{~~}
\multicolumn{1}{|p{1.18in}}{right blind spot} & 
\multicolumn{1}{|p{3.36in}|}{Over-the-shoulder, to the right} \\
\hhline{~~}
\multicolumn{1}{|p{1.18in}}{left blind spot} & 
\multicolumn{1}{|p{3.36in}|}{Over-the-shoulder, to the left} \\
\hhline{~~}
\multicolumn{1}{|p{1.18in}}{passenger} & 
\multicolumn{1}{|p{3.36in}|}{The passenger in the front passenger seat} \\
\hhline{~~}
\multicolumn{1}{|p{1.18in}}{other} & 
\multicolumn{1}{|p{3.36in}|}{All other non-road locations } \\
\hhline{--}

\end{tabular}
 \end{table}


\vspace{\baselineskip}
\subsection*{Conclusion}
\addcontentsline{toc}{subsection}{Conclusion}
It is the authors’ hope that this dataset will make a substantial contribution to computational modelling of roadway safety.\par

\subsection*{References}
\addcontentsline{toc}{subsection}{References}
\begin{adjustwidth}{0.5in}{0.0in}
International Organization for Standardization (2014). \textit{ISO 15007-1 Road vehicles—Measurement of driver visual behaviour with respect to transport information and control systems—Part 1: Definitions and parameters. }ISO, Geneva, Switzerland. \par

\end{adjustwidth}

\begin{adjustwidth}{0.5in}{0.0in}
Introducing JSON. (n.d.). Retrieved August 17, 2018, from http://www.json.org/\par

\end{adjustwidth}

\begin{adjustwidth}{0.5in}{0.0in}
Mehler, B., Kidd, D., Reimer, B., Reagan, I., Dobres, J., $\&$  McCartt, A. (2016). Multi-modal assessment of on-road demand of voice and manual phone calling and voice navigation entry across two embedded vehicle systems. \textit{Ergonomics, 59}, 344–367. \par

\end{adjustwidth}

\begin{adjustwidth}{0.5in}{0.0in}
Mehler, B., Reimer, B., Dobres, J., $\&$  Coughlin, J.F. (2015a). \textit{Assessing the Demands of Voice Based In-Vehicle Interfaces - Phase II Experiment 3 - 2015 Toyota Corolla}. Massachusetts Institute of Technology, 2015. \par

\end{adjustwidth}

\begin{adjustwidth}{0.5in}{0.0in}
Mehler, B., Reimer, B., Dobres, J., McAnulty, H., Mehler, A., Munger, D., $\&$  Coughlin, J.F. (2014a). \textit{Further Evaluation of the Effects of a Production Level $``$Voice-Command$"$  Interface on Driver Behavior: Replication and a Consideration of the Significance of Training Method.} MIT AgeLab, Cambridge, MA. \par

\end{adjustwidth}

\begin{adjustwidth}{0.5in}{0.0in}
Mehler, B., Reimer, B., Dobres, J., McAnulty, H., $\&$  Coughlin, J.F. (2015b). \textit{Assessing the Demands of Voice Based In-Vehicle Interfaces -Phase II Experiment 1 - 2014 Chevrolet Impala}. Massachusetts Institute of Technology, Cambridge, MA. \par

\end{adjustwidth}

\begin{adjustwidth}{0.5in}{0.0in}
Mehler, B., Reimer, B., McAnulty, H., Dobres, J., Lee, J., $\&$  Coughlin, J.F. (2014b). \textit{Assessing the Demands of Voice Based In-Vehicle Interfaces - Phase II Experiment 2 - 2014 Mercedes CLA}. Massachusetts Institute of Technology, Cambridge, MA, 2015. \par

\end{adjustwidth}

\begin{adjustwidth}{0.5in}{0.0in}
National Highway Traffic Safety Administration. (2012). \textit{Visual-Manual NHTSA Driver Distraction Guidelines for In-Vehicle Electronic Devices.} National Highway Traffic Safety Administration, Washington, D.C. \par

\end{adjustwidth}

\begin{adjustwidth}{0.5in}{0.0in}
Reimer, B., Mehler, B., Dobres, J., $\&$  Coughlin, J. F. (2015). \textit{Assessing the Demands of Voice Based In-Vehicle Interfaces-Phase II Experiment 4-An Exploratory Study of Driver Behavior With and Without Assistive Cruise Control (ACC)(D)}. MIT AgeLab Technical Report 2015-15. \par

\end{adjustwidth}

\begin{adjustwidth}{0.5in}{0.0in}
Sawyer, B. D., Mehler, B., $\&$  Reimer, B. (2017). Toward an Antiphony Framework for Dividing Tasks into Subtasks.\par

\end{adjustwidth}

\printbibliography
\end{document}